\documentclass{iopart}
\usepackage[T1]{fontenc}
\usepackage{graphicx,url}
\usepackage{hyperref}
\usepackage{textcomp}
\usepackage[caption=false]{subfig}
\usepackage[colorinlistoftodos]{todonotes}

\newcommand{\degree}{\ensuremath{^\circ}}
\newcommand{\sqrtHz}{\ensuremath{\sqrt{\rm Hz}}}
\newcommand{\rootHz}{\ensuremath{\sqrt{\rm Hz}}}
\newcommand{\rtHz}{\ensuremath{\sqrt{\rm Hz}}}
\usepackage{upgreek}
\begin{document}

\title{Sub-pm/$\mathrm{\mathbf{\rootHz{}}}$ non-reciprocal noise in the LISA backlink fiber}

\author{Roland Fleddermann\footnote{Current address: Department of Quantum Science, Centre for Gravitational Physics, Research School of Physics and Engineering, The Australian National University, Canberra, ACT 2601, Australia}, Christian Diekmann, Frank Steier, Michael Tr\"obs , Gerhard Heinzel and Karsten Danzmann}

\address{Albert-Einstein-Institute Hanover, Max Planck Institute for Gravitational Physics and Leibniz Universit{\"a}t Hannover, Callinstr. 38, D-30167 Hannover, Germany}

\ead{Roland.Fleddermann@anu.edu.au, Gerhard.Heinzel@aei.mpg.de}

\begin{abstract}
The future space-based gravitational wave detector Laser Interferometer Space Antenna (LISA) requires bidirectional exchange of light between its two optical benches on board of each of its three satellites. The current baseline foresees a polarization-maintaining single-mode fiber for this backlink connection.

Phase changes which are common in both directions do not enter the science measurement, but differential ("non-reciprocal") phase fluctuations directly do and must thus be guaranteed to be small enough.

We have built a setup consisting of a Zerodur\texttrademark{}  baseplate with fused silica components attached to it using hydroxide-catalysis bonding and demonstrated the reciprocity of a polarization-maintaining single-mode fiber at the 1 pm/$\sqrt{\textrm{Hz}}$ level as is required for LISA. We used balanced detection to reduce the influence of parasitic optical beams on the reciprocity measurement and a fiber length stabilization to avoid nonlinear effects in our phase measurement system (phase meter). For LISA, a different phase meter is planned to be used that does not show this nonlinearity. We corrected the influence of beam angle changes and temperature changes on the reciprocity measurement in post-processing.
\end{abstract}

\maketitle

\section{Introduction}

The baseline design of Laser Interferometer Space Antenna (LISA) \cite{Jennrich2009,LISAMissionProposal17ARXIV} foresees optical fibers to send light from one of the two optical benches on board each satellite to the other and vice versa. 

While the same laser light that is sent out to the remote satellite can be used to measure the phase of the incoming light, a different light source is foreseen to be used as local oscillator for the measurement of the distance fluctuations between the test mass and the optical bench. It is planned to use the laser for the other optical bench for this purpose. Furthermore, an independent phase reference is required to enable frequency noise subtraction using the time delay interferometry scheme \cite{Otto12CQG,Tinto2005}. Therefore, an optical connection between the two optical benches is required.

The angles between the satellites deviate periodically during one year from their nominal value of $60\degree$ by up to $1.5\degree$ while the constellation moves around the Sun \cite{Hughes2005AAS,Li2009}. Therefore, the direction of the light beams sent out from the spacecraft has to be changed. This is to be accomplished by changing the angle between two individual optical benches inside each satellite. 

While there are other proposed approaches for the task of optically connecting the two moving optical benches \cite{Weise09JPCS,Brugger2014}, a Fiber is the current base line approach . The path length noise of this fiber is subtracted by a technique called Time Delay Interferometry (TDI) \cite{Otto12CQG,Tinto2005} during data post processing. This technique requires mutual phase referencing between the two optical benches with light being sent in both directions. Phase changes which are common in both directions cancel in the TDI algorithm, but differential (``non-reciprocal'') phase fluctuations directly enter the pm/\rtHz{} science measurement and must thus be guaranteed to be small enough. Hence, we measured the non-reciprocal phase noise of such fibers in order to verify that this noise source does not spoil the measurement performance. 

A generic allocation for the non-reciprocal noise of the fiber is 1\,pm/\sqrtHz{}, which corresponds to a non-reciprocal phase noise of 6\,$\upmu$rad/$\sqrt{\textrm{Hz}}$. At this level, its contribution to the overall noise budget of 18\,pm/\sqrtHz{} \cite{LISA98,Jennrich2011} remains negligible. To measure this phase noise a setup with an inherent non-reciprocity well below this level is required.

\section{Experimental setup}

A Sagnac-interferometer \cite{ArdittyOL1981,Sagnac19132CR,Sagnac1913CR} at first glance seems ideal for the purpose of measuring non-reciprocal phase-shifts. However, a previous experiment using such a device was limited by setup noise much higher than the requirements that we aim for in this experiment \cite{Fleddermann2009JPCS}.

Therefore, we decided to build a setup resembling a representative cut-out of the situation on board the LISA satellites. In order to not be limited by thermal expansion effects, the setup was built upon a Zerodur\texttrademark{} base plate using the technique of hydroxide catalysis bonding \cite{Elliffe2005CQG}. The build process is described in detail in \cite{Steier2009CQG}.

\begin{figure}[htbp]
	\centering
		\includegraphics[width=\textwidth]{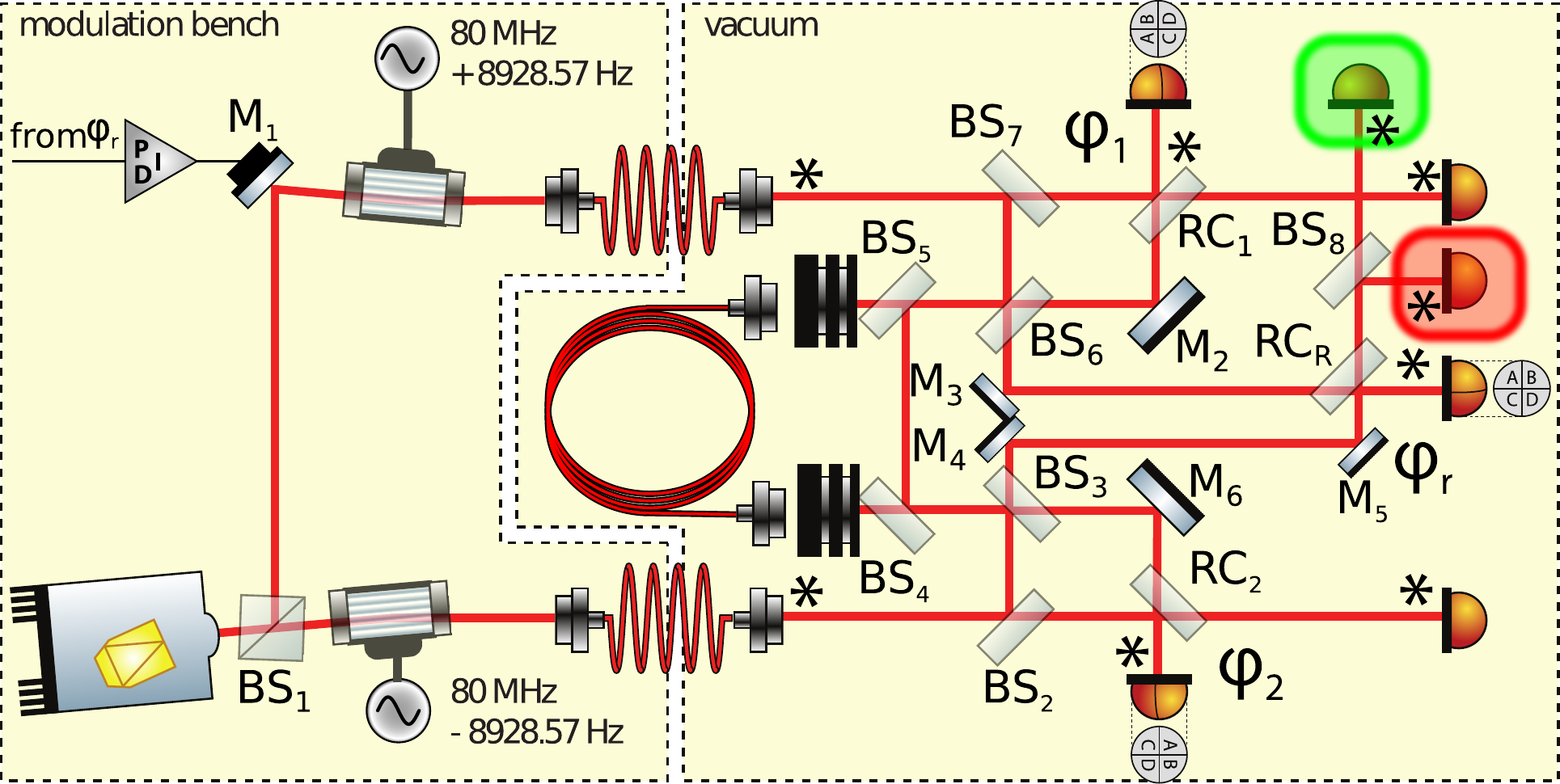}
	\caption{Measurement setup. On the modulation bench laser light is split in two parts, frequency shifted and transferred via fibers to the vacuum part of the setup. It consists of the fiber under test, two measurement and one reference interferometer. In the measurement interferometers (RC$_1$, RC$_2$), one of the interfering beams is sent through the fiber under test. In the reference interferometer (RC$_{\textrm{R}}$) the interfering beams do not pass through the fiber under test. Beam splitter BS, mirror M, recombination beam splitter RC. Asterisks (*) mark the positions where polarizers were installed.}
	\label{fig:Non-rec_test_setup_differential_PD_noise_no_polarizers}
	\label{fig:Non-rec_test_setup_polarizers}
\end{figure}

Figure \ref{fig:Non-rec_test_setup_differential_PD_noise_no_polarizers} shows a detailed schematic of the setup that was used in our experiments. On the left hand side of the figure the modulation bench is shown, where light from an iodine stabilized laser was split by a beam splitter. Both beams passed through individual AOMs, driven at 80 MHz $\pm$ 8929 Hz, resulting in two output beams with a frequency offset of 17.85~kHz.

These two beams were then coupled into fibers to transfer them into the vacuum chamber housing the main part of the experiment. This chamber is shown on the right hand side of the figure. Half of each beam was then directed towards the fiber under test and traversed it in opposite directions. The output of the fiber under test was used in an interferometer where it interfered with part of the other beam which did not travel through the fiber. Phase measurements on each of the two interferometers allowed to measure the non-reciprocal phase shift. A third interferometer was also included, where both beams interfered directly to subtract common phase shifts stemming from path length changes in the modulation bench and the fibers transporting the light into the vacuum chamber. 

Two beam splitters in front of the fiber under test (BS$_4$ and BS$_5$) allowed to bypass the fiber and measure the non-reciprocal noise of the test setup to find its limiting noise floor. Two photo diodes which sensed part of the input beams power allowed for an active laser power stabilization.

\begin{figure}[htbp]
	\centering
	\includegraphics[width=0.75\textwidth]{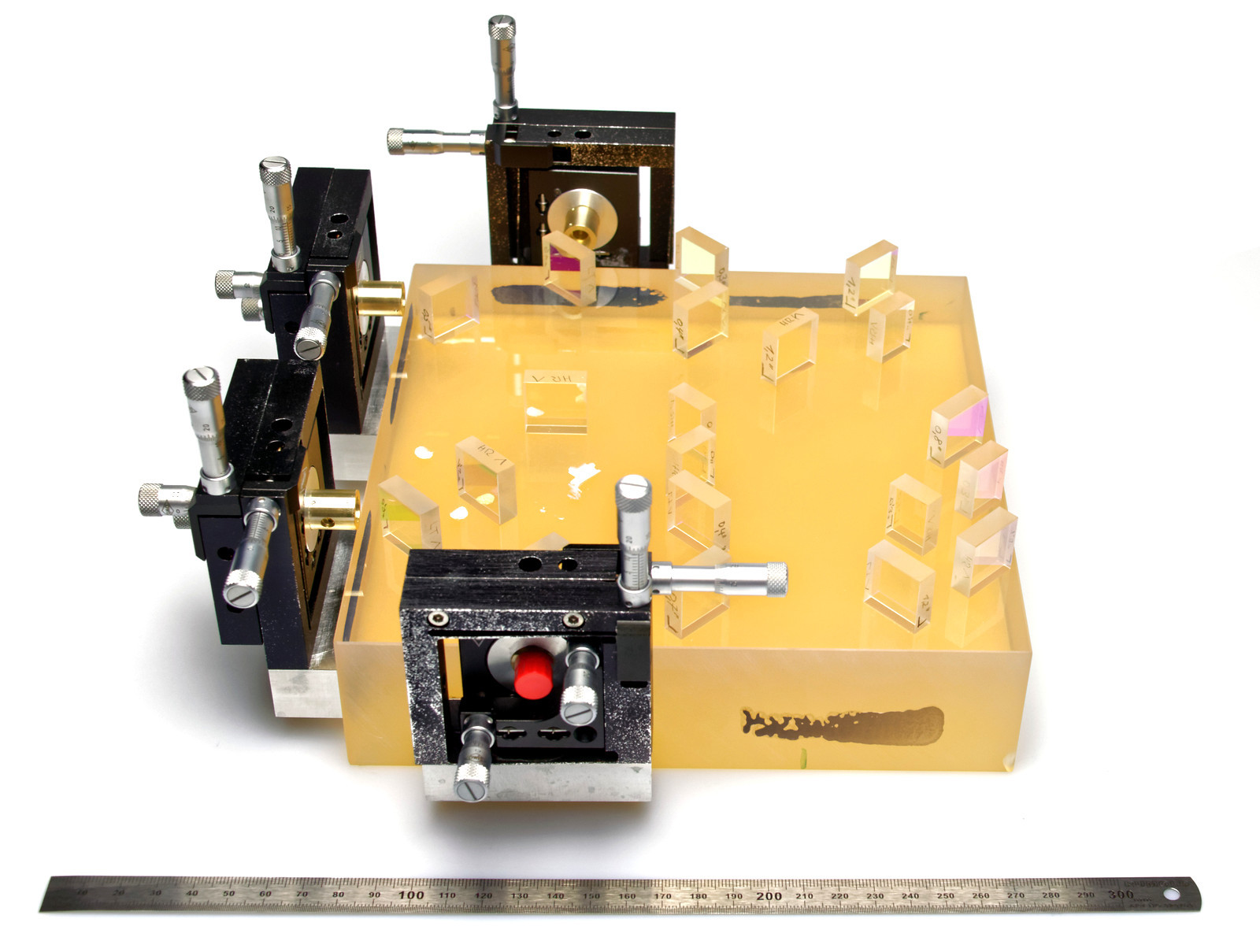}
	\caption{Photograph of the stable interferometer with bonded optical components and fiber injectors for the fiber under test (on the left) and for the delivery fibers. The photo diodes reading out the interferometers are not shown. They were placed next to the Zerodur\texttrademark{}  baseplate.}
	\label{fig:Backlink-Ifo}
\end{figure}

Figure~\ref{fig:Backlink-Ifo} shows a photograph of the stable interferometer with bonded optical components and fiber injectors for the fiber under test (on the left) and for the delivery fibers. The photo diodes reading out the interferometers are not shown. They were placed next to the Zerodur\texttrademark{} baseplate. The interferometer pictured contains the components shown in the ``vaccum'' section of Figure~\ref{fig:Non-rec_test_setup_polarizers}. For details on the build process and placement of the optical and auxiliary components refer to \cite{Steier2009CQG}.

\section{Measurement principle}

In the reference interferometer ($\varphi_{\rm{R}}$), illustrated in Figure~\ref{fig:Non-rec_test_setup_differential_PD_noise_no_polarizers}, both beams were directly routed to the recombination beam splitter RC$_{\rm{R}}$ without passing through the fiber. The measured phase of this interferometer was used for subtraction of the path length changes on the modulation bench which are common-mode in all interferometers. Using the light wavelength $\lambda$ its phase is hence found to be 

\begin{equation}
\varphi_\mathrm{r}=\frac{2\pi}{\lambda}\left(\overline{\mathrm{BS_1BS_2BS_3M_4M_5RC_R}}-\overline{\mathrm{BS_1M_1BS_7BS_6M_3RC_R}}\right)
\end{equation}

In the first measurement interferometer ($\varphi_{\rm{1}}$) the first beam was traveling directly to the interference beam splitter, while the second beam passed through the fiber on its way to this point. The situation in the second measurement interferometer ($\varphi_{\rm{2}}$) was exactly opposite. Here, the second beam was traveling directly to the beam splitter, while the first beam passed through the fiber. The phase in these two interferometers is hence found to be 

\begin{equation}
\varphi_\mathrm{1} = \frac{2\pi}{\lambda}\left(\overline{\mathrm{BS_1BS_2BS_3BS_4BS_5BS_6M_2RC_1}}-\overline{\mathrm{BS_1M_1BS_7RC_1}} \right)
\end{equation}
\begin{equation}
\varphi_\mathrm{2} = \frac{2\pi}{\lambda}\left(\overline{\mathrm{BS_1BS_2RC_2}}-\overline{\mathrm{BS_1M_1BS_7BS_6BS_5BS_4BS_3M_6RC_2}}\right)
\end{equation}

% s1 = BS1 BS2 BS3 BS4 BS5 BS6 M2 RC1 - BS1 M1 BS7 RC1
% s2 = BS1 BS2 RC2 - BS1 M1 BS7 BS6 BS5 BS4 BS3 M6 RC2
% sr = BS1 BS2 BS3 M4 M5 RCR -  BS1 M1 BS7 BS6 M3 RCR

% s1+s2 = BS3 BS4 BS5 BS6 - BS6 BS5 BS4 BS3 + BS1 BS2 BS3 + BS6 M2 RCR1 + BS1 BS2 RC2 - BS1 M1 BS7 RC1 - BS1 M1 BS7 BS6  - BS3 M6 RC2

% s1+s2-2sr = BS3 BS4 BS5 BS6 - BS6 BS5 BS4 BS3 + 2 BS6 M3 RCR - 2 BS3 M4 M5 RCR + BS6 M2 RC1 + BS2 RC2 - BS3 M6 RC2 - BS7 RC1 - BS2 BS3  + BS7 BS6

Thus the reciprocal phase changes were measured in the two measurement interferometers ($\varphi_{\rm{1}}$ and $\varphi_{\rm{2}}$) in opposite directions. The non-reciprocity was found by adding the changes in the phases measured in both interferometers. 

\begin{eqnarray}
\label{eq:non_rec1}
\varphi_\mathrm{1}+\varphi_\mathrm{2} &= \frac{2\pi}{\lambda}(\underbrace{(\overline{\mathrm{BS_3BS_4BS_5BS_6}}-\overline{\mathrm{BS_6BS_5BS_4BS_3}})}_{\mathrm{non-reciprocity}}\nonumber
\\
&+\overline{\underbrace{\mathrm{BS_1BS_2}}_\mathrm{unstable}\mathrm{BS_3}}+\overline{\mathrm{BS_6M_2RC_1}}+\overline{\underbrace{\mathrm{BS_1BS_2}}_\mathrm{unstable}\mathrm{RC_2}}\nonumber
\\
&-\overline{\underbrace{\mathrm{BS_1M_1BS_7}}_\mathrm{unstable}\mathrm{RC_1}}-\overline{\underbrace{\mathrm{BS_1M_1BS_7}}_\mathrm{unstable}\mathrm{BS_6}}-\overline{\mathrm{BS_3M_6RC_2}})
\end{eqnarray}

However, as can be seen from the above equation, path length changes in the modulation bench (left half of Figure~\ref{fig:Non-rec_test_setup_differential_PD_noise_no_polarizers}) will still be present in this combination of signals. Since these paths contain long optical fibers, their phase is expected to change by many orders of magnitude more than can be tolerated in this experiment. This can be overcome by subtracting the reference interferometer's phase $\varphi_r$ from each of these signals. 

Looking at the path length differences above, we find that if one  subtracts the reference phase $\varphi_r$ twice from equation (\ref{eq:non_rec1}) the result is:
\begin{eqnarray} 
\fl\varphi_1 + \varphi_2 - 2 \varphi_r =& \frac{2\pi}{\lambda}(\underbrace{(\overline{\mathrm{BS_3BS_4BS_5BS_6}}-\overline{\mathrm{BS_6BS_5BS_4BS_3}})}_{\mathrm{non-reciprocity}}\nonumber
\\
&+\underbrace{2\overline{\mathrm{BS_6M_3RC_R}}-2\overline{\mathrm{BS_3M_4M_5RC_R}}+\overline{\mathrm{BS_7BS_6}}-\overline{\mathrm{BS_2BS_3}}}_\mathrm{stable}\nonumber
\\
&+\underbrace{\overline{\mathrm{BS_6M_2RC_1}}+\overline{\mathrm{BS_2RC_2}}-\overline{\mathrm{BS_3M_6RC_2}}-\overline{\mathrm{BS_7RC_1}}}_\mathrm{stable}
\end{eqnarray}

As the path lengths on the Zerodur\texttrademark{}  optical bench are very stable, this sum can be considered an upper limit on the non-reciprocal path length noise.

\section{Setup Characterization}
\label{sec:Setup_Characterization}

\subsection{Phase measurement system}
\label{sec:phasemeter_noise}
In the measurement scheme presented in the previous section, the path length changes were converted to phase changes of the optical signals, which were in turn converted to changes in the phase of the heterodyne interference signal at the heterodyne frequency of 17.85~kHz. The light was detected by photo diodes at the corresponding beam splitters and the phase of the resulting electrical signals was measured with a single-bin Fourier transform phase meter implemented on a field programmable gate array \cite{Heinzel2004CQG}, which yields the phase and amplitude of the heterodyne signal. 

Tests were performed to assess the phase measurement noise of the phase meter. An electrical signal from a signal generator was split into two equal parts and sensed by two channels of the phase meter. The difference of the phases in both channels was then computed. This yielded a lower limit on the phase measurement noise to be expected. 

\begin{figure}
	\centering
		\includegraphics[width=\linewidth]{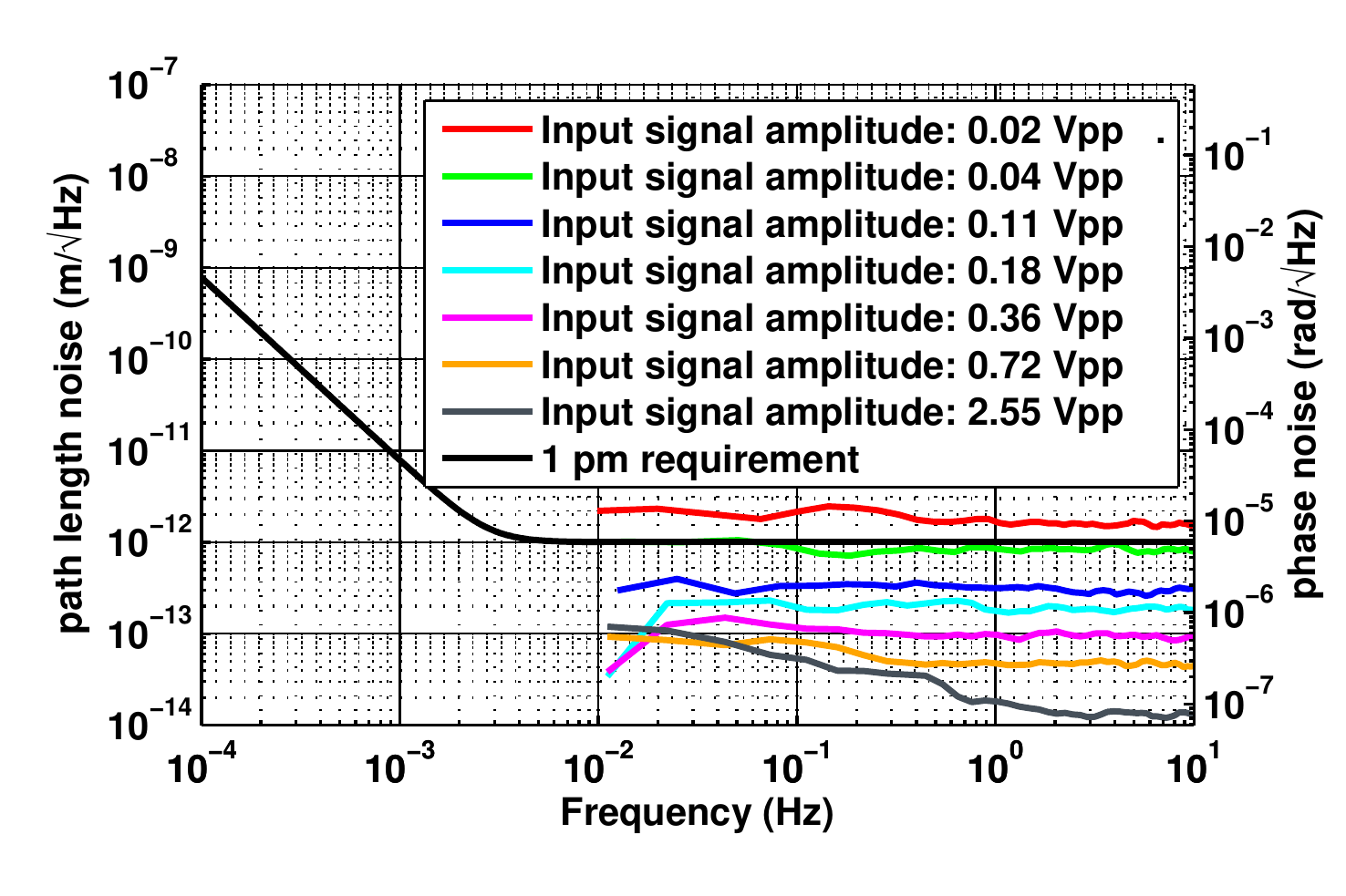}
	\caption{Phase measurement noise using different input signal levels from a signal generator. The signal from the signal generator was split into two equal parts and sensed by two channels of the phase meter. The difference of the phases in both channels was then computed.}
	\label{fig:phasemeter_noise_input_levels_spectra}
\end{figure}

The results of these measurements are presented in Figure \ref{fig:phasemeter_noise_input_levels_spectra}. Several noise spectral densities are shown, each representing the differential phase noise between two channels being driven from a common signal generator signal. The amplitude of this input signal was varied from 20~mV$_\mathrm{pp}$ to 2.5~V$_\mathrm{pp}$ between the different measurements. As can be seen, the phase measurement noise was compliant with the requirement of a 1 pm/\rootHz{} measurement as long as the input signal amplitude remained above approximately 40~mV$_\mathrm{pp}$. This was the case in all actual optical experiments conducted. In fact, typical input signals exceeded 1~V$_\mathrm{pp}$ in amplitude.

\subsubsection{Differential noise between photo diodes}

In the previous Section \ref{sec:phasemeter_noise} the phase meter noise using electrical input signals was measured. This gave only a lower limit of the phase meter noise. In the real experimental setup the signals are not perfect sine waves with a constant amplitude, but instead they may have distortions and amplitude noise. Therefore it is necessary to test the ability of the phase meter to measure the phase of such realistic signals. 

In order to measure this noise, the setup was slightly changed to include an additional photo diode. A schematic of this modified setup can be found in Figure~\ref{fig:Non-rec_test_setup_differential_PD_noise_no_polarizers}. The additional photo diode is highlighted by a red frame in the right hand part of the figure. This photo diode was used to measure the phase of a beam that was split off one of the two reference interferometer outputs. The phase observed at this point should exactly match the phase observed at the original interferometer output found in the upper right part of the Figure, highlighted in a green frame. This allowed us to calculate the difference between the two phases observed by these two photo diodes and to thus get a measure of the so-called differential photo diode noise.

\begin{figure}[htbp]
	\centering
		\includegraphics[width=\columnwidth]{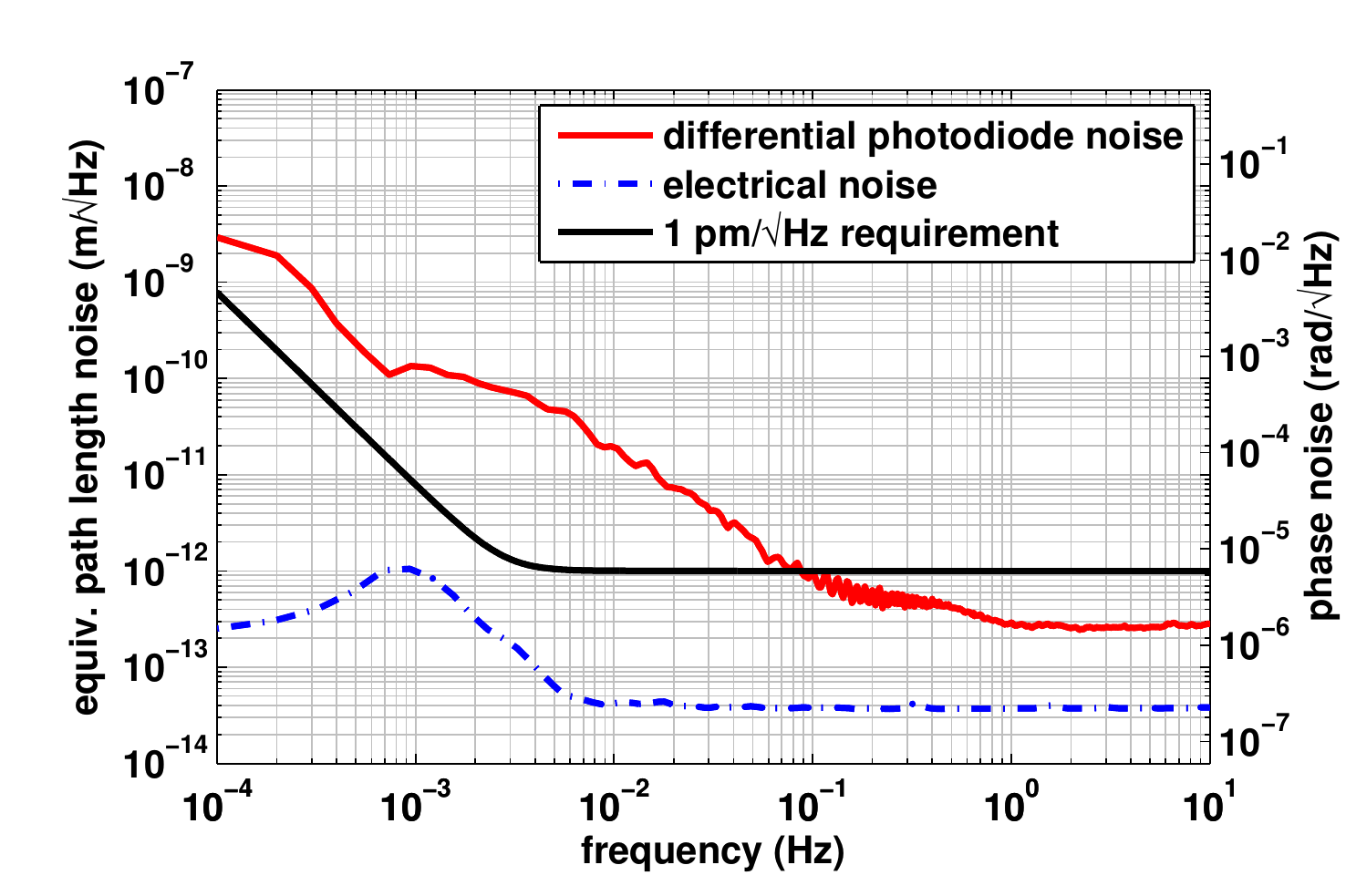}
	\caption{Differential photo diode noise compared to the electrical phase meter noise. The signals from the photo diodes marked with a red and a green frame in Figure~\ref{fig:Non-rec_test_setup_differential_PD_noise_no_polarizers} were input to the phase meter and their difference phase was computed. Although the photo diode signals should be identical, significantly higher noise was measured than in the electrical test of the phase meter.}
	\label{fig:differential_pd_noise_initial_compared_to_pm_noise}
\end{figure}

First measurements using this setup showed that the differential noise found between the signals from the two photo diodes sensing the same split beam was substantially higher than differential noise between two phase meter channels sampling the same electrically fed signal. The respective noise spectral density plots can be found in Figure \ref{fig:differential_pd_noise_initial_compared_to_pm_noise}. Here, the red trace represents the differential photo diode noise. This noise is at 0.3 pm/\rootHz{} for frequencies above one Hertz, increasing roughly as 1/f towards lower frequencies. It violates the 1\,pm requirement at frequencies below 0.1\,Hz. Here, the electronic noise of the phase meter, as measured previously, is represented by the blue trace. This noise has an amplitude of about 0.05 pm/\rootHz{} for frequencies above 5\,mHz. It is also well below the 1\,pm requirement in the whole measurement band. From this one can conclude that additional noise is present in this differential photo diode measurement.

\subsubsection{Polarizers in front of photo diodes}
\label{sec:differential_pd_noise}	
A potential source of differential photo diode noise is a spurious interferometer formed with the light that is not in the nominal s-polarization, but in the orthogonal p-polarization state. This light could stem from imperfect polarization maintaining properties of the fibers or it might be introduced through scattering processes. 

To mitigate the effect of this spurious interferometer, polarization filters were introduced in front of the photo diodes in order to reject the light from this spurious interferometer. Polarcor\textsuperscript{\textregistered}{} filters with an extinction ratio of up to $10^{6}$ were used to enable high rejection of unwanted signals. The measurements with these filters installed were then compared to previous measurements without polarization filters.

\begin{figure}[htbp]
	\centering
		\includegraphics[width=\columnwidth]{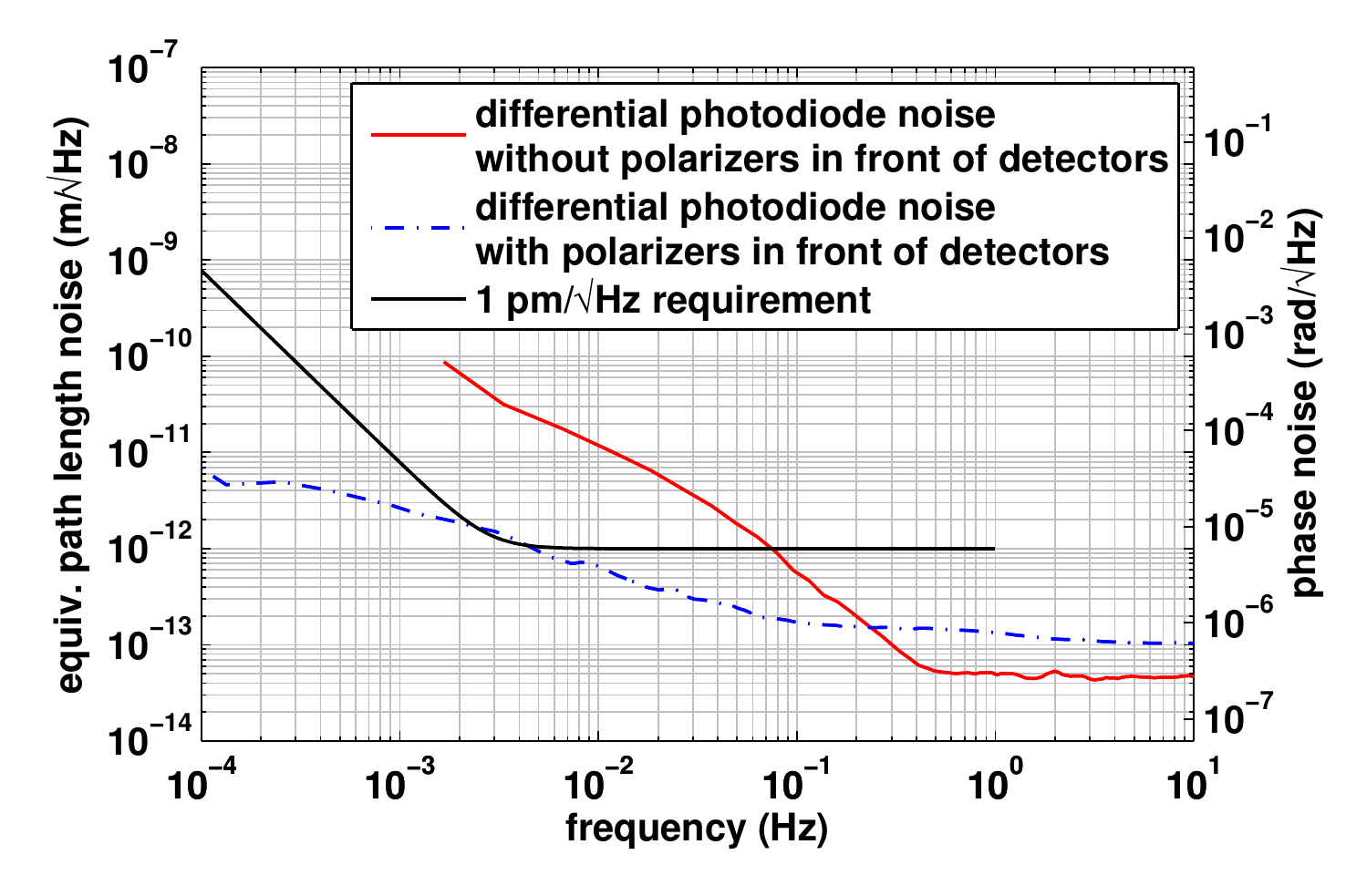}
	\caption{Differential photo diode noise with and without polarizers in front of  the photo diodes marked with a red and a green frame in Figure~\ref{fig:Non-rec_test_setup_differential_PD_noise_no_polarizers}. Polarizers significantly reduced the noise.}
	\label{fig:compare_pol_no_pol_2}
\end{figure}

The resulting noise spectral densities are presented in Figure~\ref{fig:compare_pol_no_pol_2}. In this figure, the blue trace representing the measurement with polarizers shows higher noise than found in the measurement without polarizers represented by the red trace, for frequencies above 0.2 Hz. This can be attributed to the signal loss occurring due to the transmission of about 70\% at 1064\,nm wavelength of the polarization filters. However, the noise below 0.2~Hz is lower with the polarization filters in place. It is compatible with the 1\,pm/\rootHz{} requirement in the whole frequency range of interest (0.1~mHz to 1~Hz).
	
\subsection{Optical path length difference stabilization}
\label{sec:OPD}
\label{sec:opd}
Previous experiments using this phase measurement setup had identified a non-linearity in the measurement process arising from interfering electronic signals \cite{Wand2006CQG}. In the experiments described therein, the non-linearity could be overcome through stabilization of the differential path length changes of the vacuum feed-through fibers bringing the light to the experiment. 

The same effect also disturbed measurements in this experiment, because the optical path length difference (OPD) between the two beams generated on the modulation bench was not stabilized. While this difference normally should not cause any problems because it is common to all measurement signals and cancels in the subtraction of the reference interferometer's phase, it still posed a problem in this experiment due to non-linear effects in the phase measurement system. Measurements in the different interferometers no longer showed exactly the same common phase shift, but depended on the phase measured in each interferometer.

\begin{figure}
	\centering
		\includegraphics[width=\columnwidth]{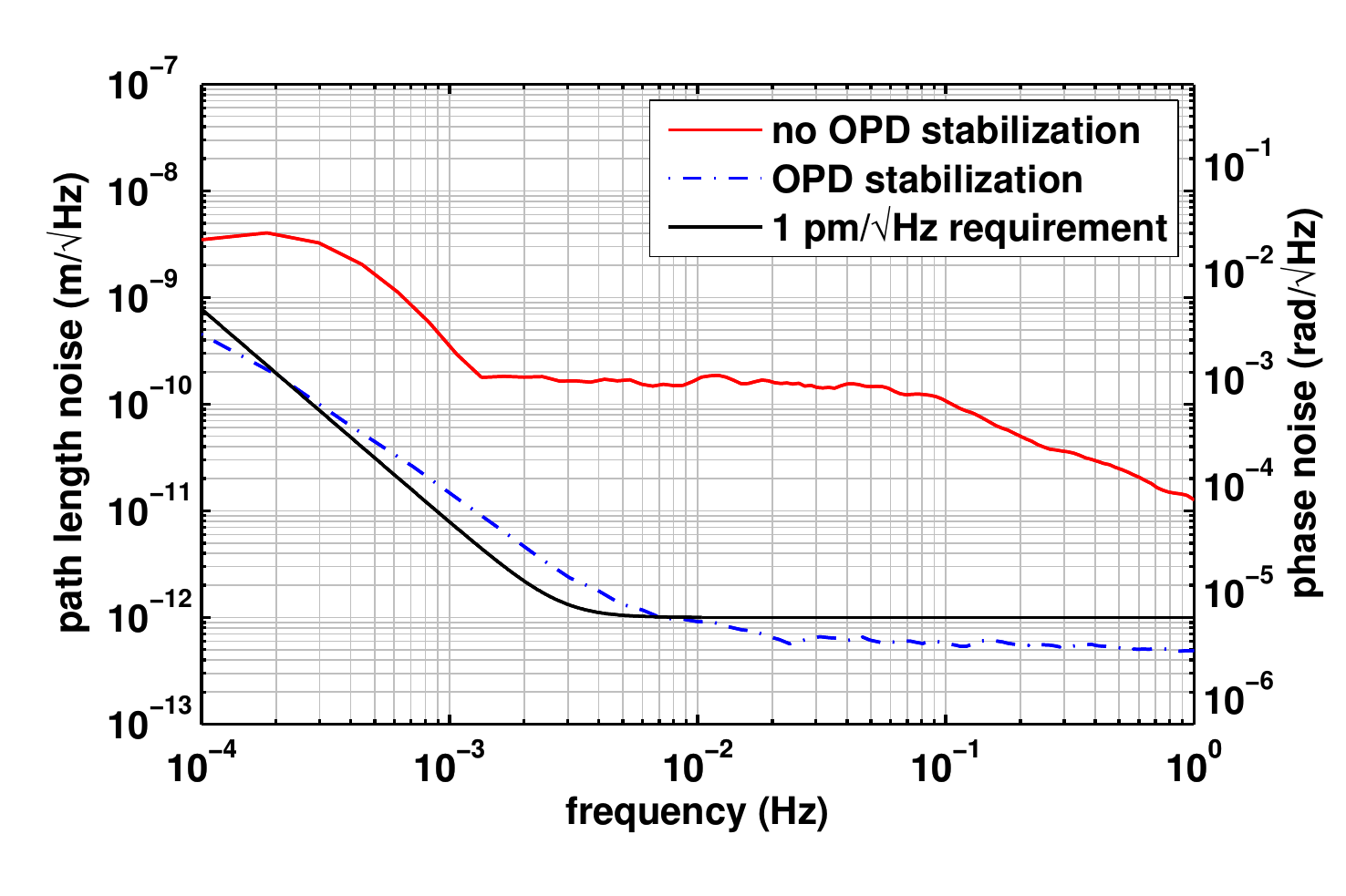}
	\caption{Comparison between non-reciprocal noise observed with active optical path length difference (OPD) stabilization and without.}
	\label{fig:compare_opd_no_opd}
\end{figure}

Measurements of the non-reciprocity \--- in fact any interferometric length measurements using this phase measurement system \--- without active stabilization of the differential path length consequently show a typical shoulder-shaped noise curve. A detailed analysis is given in \cite{Wand2006CQG}.
Figure \ref{fig:compare_opd_no_opd} shows an example of such a noise curve observed in the measurement of the non-reciprocity of the setup. The red trace without stabilization shows a noise level of about 200\,pm/\rootHz{} between 1\,mHz and 0.1\,Hz, rolling off with $1/f$ toward higher frequencies. The white noise level between 1\,mHz and 0.1\,Hz is caused by the non-linear effects in the phase measurement process, leading to a peak-to-peak phase variation corresponding to 200\,pm/\rootHz{}, which is spread about a wide frequency range.

To overcome this problem, OPD stabilization was integrated into the modulation bench. One beam was reflected by a piezo mounted mirror which was driven by a feedback circuit stabilizing the phase in the reference interferometer to an electronic reference phase generated by the offset-locking electronics used to drive the AOMs. The effect on the measurement sensitivity is also illustrated in Figure \ref{fig:compare_opd_no_opd}. The blue dashed trace represents the non-reciprocal noise of the setup while the OPD stabilization was active. As one can see, the observed non-reciprocal noise is at a level of 0.5~pm/\rootHz{} at high frequencies and increases as $1/f^2$ for frequencies below 5~mHz, thus following the shape of the requirement, shown in black. 

The OPD stabilization was therefore used in all subsequent measurements, successfully reducing the influence of non-linear effects in the phase measurement system to the levels required.

\subsection{Laser power stabilization}
\label{sec:laserintensity}

Another factor that could potentially impact phase measurements and thereby the observed non-reciprocal noise in the experiment is laser power noise. One potential coupling mechanism of low frequency intensity variations is a change of photo diode capacitance with temperature, which is in turn influenced by the absorbed light power. 

To assess the influence of this potential noise source measurements were performed while the light power was modulated using the amplitude modulation input of the AOM drivers. These inputs allow one to easily change the amplitude of the RF-signal driving the AOMs. This will lead to a change in the light power in the first order diffracted beam output due to the change in diffraction efficiency at different RF-signal levels. 

Using the intensity coupling coefficient obtained in these measurements, it was possible to measure the DC relative intensity fluctuations and make a projection of their influence on the non-reciprocal noise. The measured DC relative intensity noise (RIN) was calculated and multiplied with the known coupling factor which was assumed to be frequency-independent for simplicity to obtain the expected non-reciprocal noise from RIN fluctuations.

\begin{figure}
	\centering
		\includegraphics[width=\columnwidth]{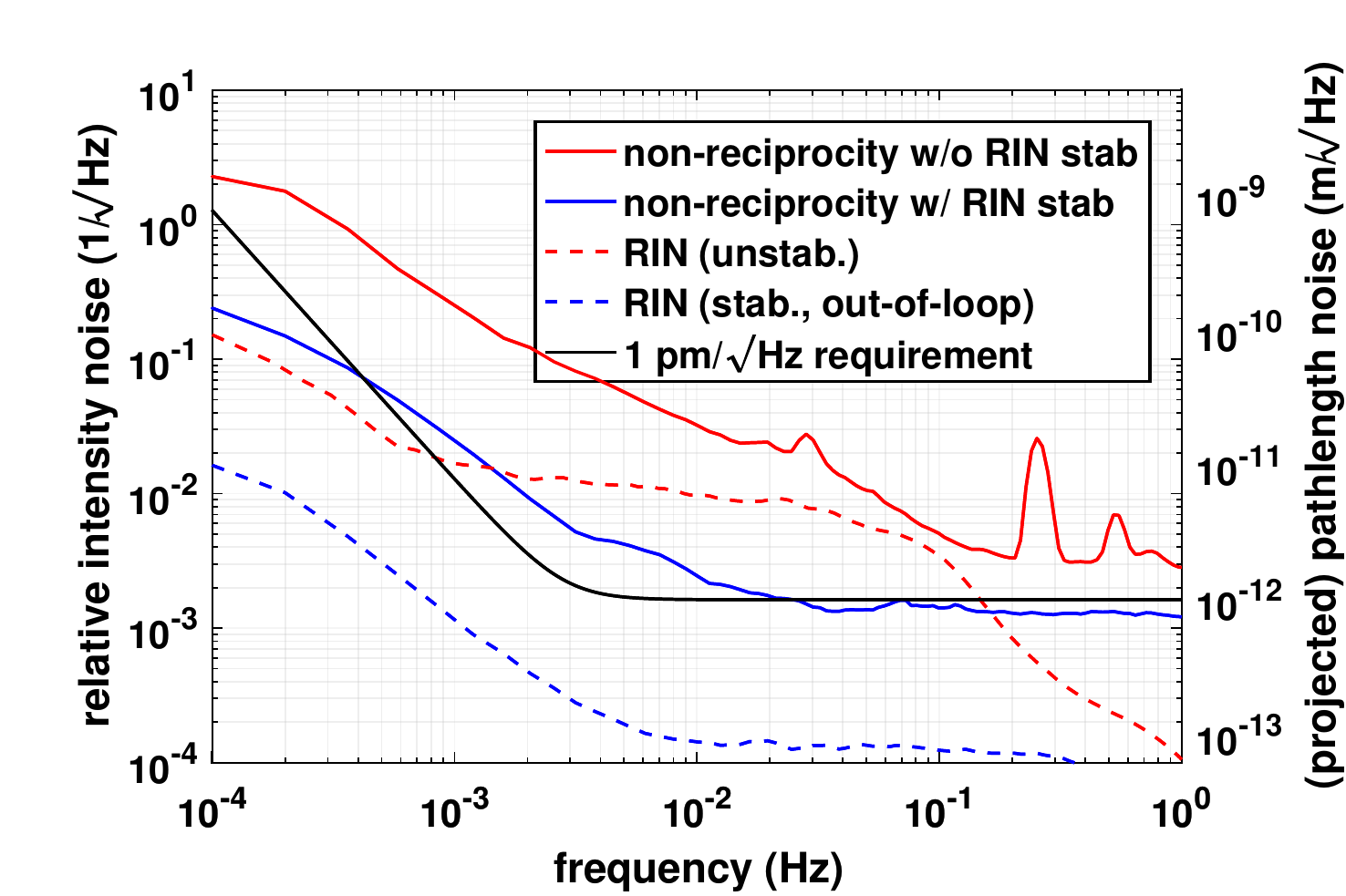}
	\caption{Comparison between non-reciprocity observed with active laser amplitude stabilization and without. The solid red (blue) trace shows the measured non-reciprocity without (with) relative intensity noise (RIN) stabilization. The dashed red (blue) trace shows the measured RIN without (with) DC stabilization. The right axis shows the projected path length noise from the measured relative intensity noise.}
	\label{fig:compare_DC_stab_no_DC_nonrec}
\end{figure}

The results of this noise projection can be found in Figure \ref{fig:compare_DC_stab_no_DC_nonrec}. Measurements with and without active RIN stabilization are compared. As can be seen from the figure, without active RIN stabilization, the observed non-reciprocal noise is higher in the whole frequency range. The noise shape is very similar to $1/f$, starting out at about 100~pm/\rtHz{} at a frequency of 1\,mHz and diminishing toward higher frequencies. The RIN without active stabilization, shown in red and dashed,  is at about $10^{-2}$/\rtHz{} for frequencies between 2\,mHz and 0.1\,Hz. The resulting projected non-reciprocal noise is very close to the observed non-reciprocal noise at a frequency of 0.1\,Hz. With active RIN stabilization, the relative intensity noise, shown dashed and in blue, is reduced by about a factor of 100 in the frequency range between 2\,mHz and 0.1\,Hz. The solid blue trace, representing the corresponding observed non-reciprocal noise is very close to the 1\,pm requirement, plotted in black.

Therefore, one can see that observed non-reciprocal noise is greatly reduced when active laser relative intensity noise stabilization is employed. A relative intensity stability of better than $10^{-3}$ is required given the amplitude coupling factor found here to allow the measurement of non-reciprocal path length noise at a level compatible with the 1\,pm requirement.

To overcome the influence of the changes in laser intensity, an electronic intensity stabilization circuit was used. This circuit was already available from other LISA Pathfinder related experiments conducted at the AEI. It uses the photo current of an auxiliary photo diode on the modulation bench as input,
 which is passed through a trans-impedance amplifier to convert it into a proportional voltage. The resulting signal is then compared to a 10\,V voltage obtained from an AD587 voltage reference, which was found to deliver particularly low voltage noise in \cite{Fleddermann2009a} and the difference serves as error signal for a feedback loop, consisting of two integrators and a proportional amplifier. The feedback signal was then applied to the AOM drivers' amplitude modulation input.

\subsection{Laser frequency noise}
\label{sec:laserFrequencyNoise1}

In a Mach-Zehnder interferometer with unequal arms, laser frequency noise fluctuations lead to phase fluctuations in the interference signal. The relationship between frequency noise and phase noise is as follows \cite{Heinzel2004CQG}:
$$ \frac{\delta \varphi}{\delta f} = \frac{2\pi\Delta L}{c}$$
Here, $\delta \varphi$ stands for the phase induced by the frequency noise, $\delta f$. The arm length difference is denoted by $\Delta L$, and $c$ is the speed of light. 

\begin{figure}
	\centering
		\includegraphics[width=\columnwidth]{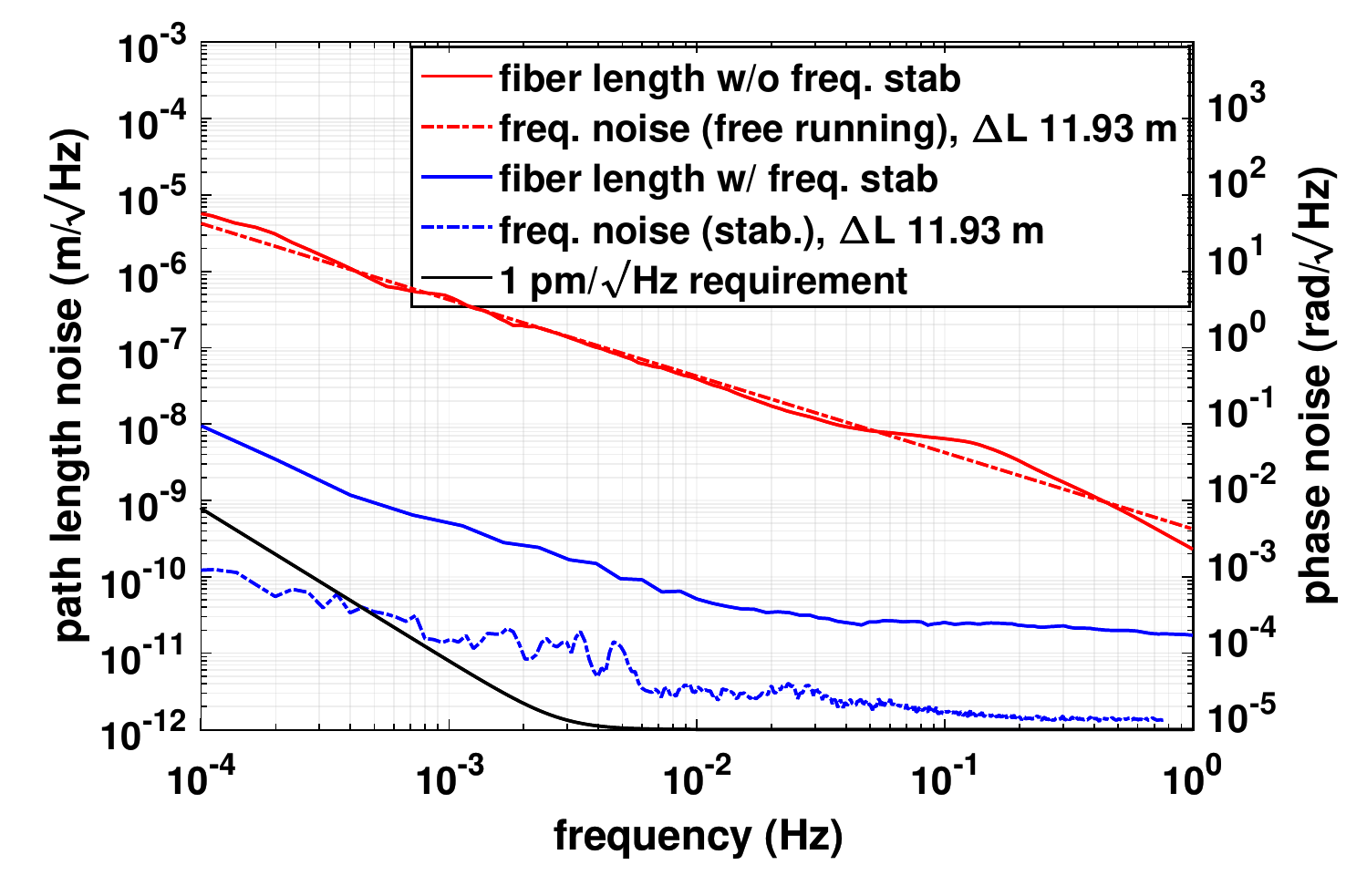}
	\caption{Measured path length noise difference between one measurement interferometer and the reference interferometer. Solid red trace: without laser frequency stabilization, solid blue trace: with laser frequency stabilization, Without laser frequency stabilization, the measurement is dominated by laser frequency noise. The dashed blue trace shows the projection of laser frequency noise on the fiber length measurement with active laser frequency stabilization.}
	\label{fig:frequency_noise_length_meas}
\end{figure}

In order to find the frequency noise coupling coefficient in the experiment, we switched off the frequency stabilization of the laser and measured the length noise of the fiber by using the phase in just one of the measurement interferometers, subtracting the reference phase. After computing the spectral density of this length noise, we found a much higher length noise than with active laser frequency stabilization. The shape of the noise curve followed a $1/f$ curve as shown in the red trace in Figure \ref{fig:frequency_noise_length_meas}. From the fact that the observed length noise increased after switching off the frequency stabilization it can be deduced that noise levels without laser frequency stabilization are dominated by this noise source.

This, in turn, allows one to compute the coupling factor between the known laser frequency noise without active stabilization of 10\,kHz$/f$ with Fourier frequency $f$ (\cite{Troebs09JOSAB}) and the observed path length noise. In this case, this yielded a coupling factor of 0.25\,\textmu rad/Hz, corresponding to a length mismatch in the arms of about 12\,m. This is in good agreement with the \emph{optical} length of the fiber plus the additional path length mismatches on the optical bench present in each individual interferometer. 

After repeating the measurement with frequency stabilization enabled, a much lower length noise was found, as indicated by the blue trace in Figure \ref{fig:frequency_noise_length_meas}. Using the frequency noise of the stabilized laser it was also possible to predict the frequency noise induced non-reciprocal path length noise. 
The frequency noise of the iodine stabilized laser was measured by comparing its frequency to that of a reference laser which was locked to an external ultra-stable reference cavity. The setup is described in \cite{Troebs09JOSAB}. The laser frequency noise induced non-reciprocal noise is represented by the dashed blue trace in the figure. 
 Because this trace is significantly below the solid blue trace, it can be concluded that the length measurement noise is no longer dominated by frequency noise induced fluctuations when active laser frequency stabilization is employed.

\begin{figure}
	\centering
		\includegraphics[width=\columnwidth]{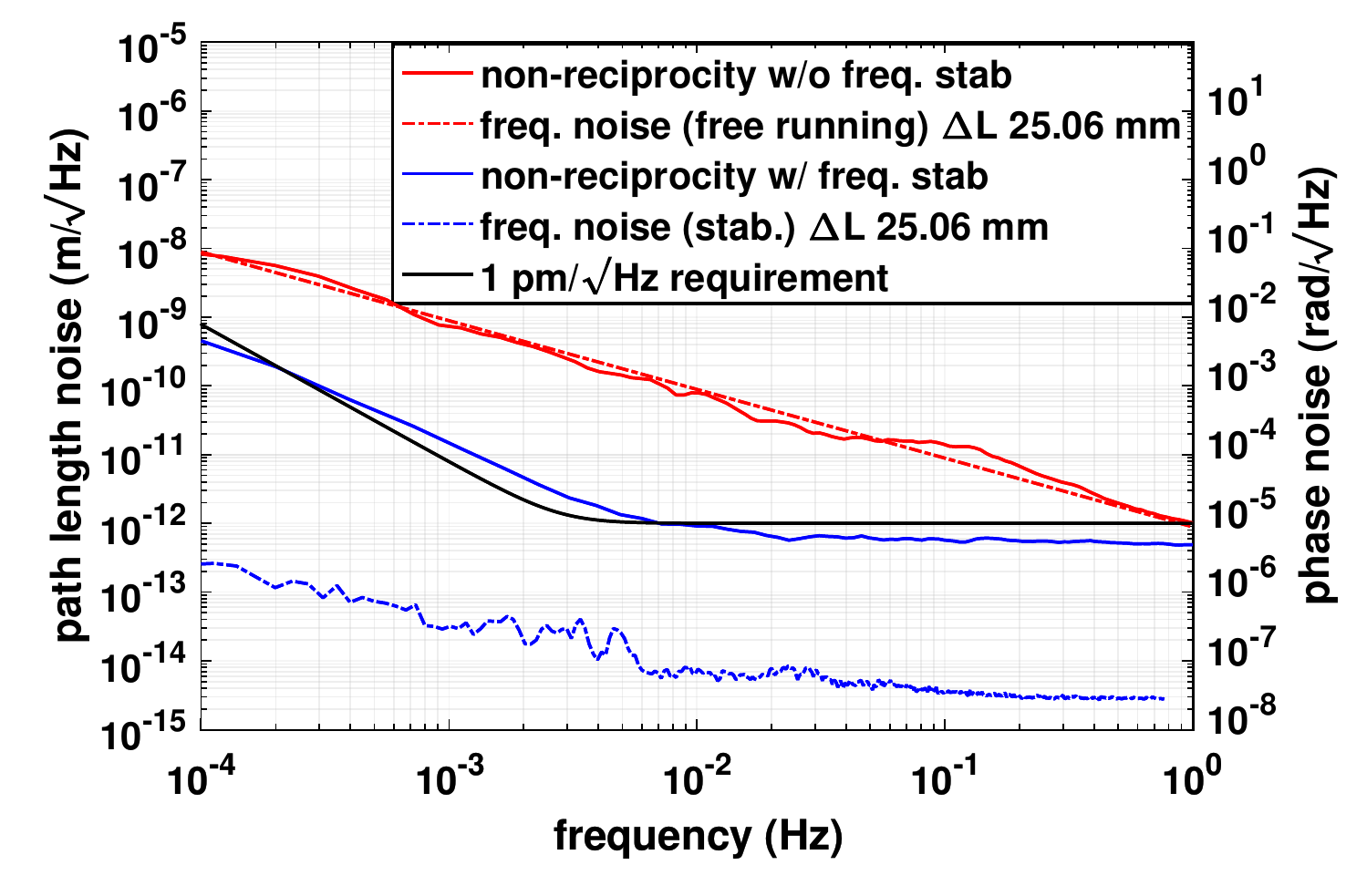}
	\caption{Non-reciprocity measurement. Solid red trace: without laser frequency stabilization, solid blue trace: with laser frequency stabilization, dash-dotted traces: projection of laser frequency noise on non-reciprocity. With laser frequency stabilization, the non-reciprocity measurement is not dominated by laser frequency noise.}
	\label{fig:frequency_noise_nonrec}
\end{figure}

The same analysis was also performed for the non-reciprocity signal combination. Again, the red trace in Figure \ref{fig:frequency_noise_nonrec} represents the non-reciprocal noise observed when the laser's frequency stabilization is switched off. In the combination of interferometer signals that yields the non-reciprocal length, the path length difference in the two measurement interferometers cancels to a large extent, and a much smaller observed frequency coupling coefficient of only approximately 0.5\,nrad/Hz results, displayed in the dashed red trace. This corresponds to an effective arm length difference of only approximately 25\,mm.

The blue trace in Figure \ref{fig:frequency_noise_nonrec} shows the observed non-reciprocal noise when the laser frequency stabilization is active. Again, the dashed blue trace shows the projected laser frequency noise for comparison. Here, from the fact that this trace is significantly below the observed non-reciprocal noise, it can also be concluded that it is no longer dominated by frequency noise induced phase fluctuations, when active laser frequency stabilization is used. However, this also means that an active frequency noise stabilization is required in order to achieve sub-picometer non-reciprocal noise measurements. Therefore, all measurements were conducted using the frequency stabilized laser.

\subsection{Polarization control}
\label{ft_sec:pol_control}

As a consequence of the measurements presented in Section \ref{sec:differential_pd_noise}, linear Polarcor\textsuperscript{\textregistered}{} polarizers were added in front of all photo diodes. The photo diode mounts were upgraded during this step to accommodate the new filters. 

Additional polarizers were also added directly behind the fiber launchers that bring the light onto the optical bench. The Polarcor\textsuperscript{\textregistered}{} filters could not be used in this position due to their insufficient wavefront quality. Instead, Glan-Thompson polarizers by Bernhard Halle Nachfl. GmbH were used which have a similar extinction ratio to the Polarcor\textsuperscript{\textregistered}{} filters but better wave front quality. However, they are more bulky and therefore inadequate for use in front of all photo diodes. The position of all polarizers in the experiment is indicated by an asterisk (*) in Figure~\ref{fig:Non-rec_test_setup_polarizers}. 

To further improve the polarization stability in the experiment, the input polarization state to both the input fibers and the fiber under test was matched to the polarization-maintaining axes. The polarization-maintaining fibers that were used in the experiment are of the PANDA-type \cite{Sasaki1983}. This type of fiber features an inner core, which guides the light, and two \emph{stress-inducing elements}, which serve to introduce stress at the fiber core in well-defined directions, which leads to stress-induced birefringence in the fiber core. 

The alignment was conducted with the help of a polarimeter by Sch{\"a}fter und Kirchhoff. 

\subsection{Fiber length stabilization}
\label{sec:fiber_length_stab}

This effect stems from electrical interference and was observed previously in measurements using the same phase meter \cite{Wand2006CQG}. It is the main reason why the optical path length difference stabilization between the two delivery fibers needed to be implemented, as described in Section \ref{sec:opd}. 

A similar problem occurs if the fiber under test expands or contracts due to the influence of temperature fluctuations. While the phases observed at both fiber ends are nominally reversed an apparent phase noise can be introduced into the non-reciprocal signal combination if the individual measurement phases change by amounts of order radians or larger due to the non-linearity of the phase meter, caused mainly by electrical interference.

By using a fiber wound on a cylindrical piezo as fiber under test it was possible to observe a periodic error in the non-reciprocal signal. When scanning the fiber length over several micrometers by applying a triangular voltage to the piezo on which it was wound, a sinusoidal response in the non-reciprocal length measurement was observed. 

To overcome this limitation, fiber length stabilization was implemented using a ring piezo. The fiber under test was wound around the piezo several times, enabling its length to be changed by applying an appropriate voltage to the ring piezo. An analog output signal of one of the measurement photo diodes % MT: photo added
 was used as input to a servo loop identical to the one used in the optical path length difference stabilization, only this time feeding back to the fiber length instead of the piezo in the modulation bench. The result is a stabilized phase readout in one of the measurement interferometers. In the case of perfect reciprocity, this would also lead to a stabilized phase in the other measurement interferometer. Residual non-reciprocity is still captured in the same way as before, combining measurements from all interferometers, only with the effect of non-linearity in the phase meter removed.

\begin{figure}
	\centering
		\includegraphics[width=\columnwidth]{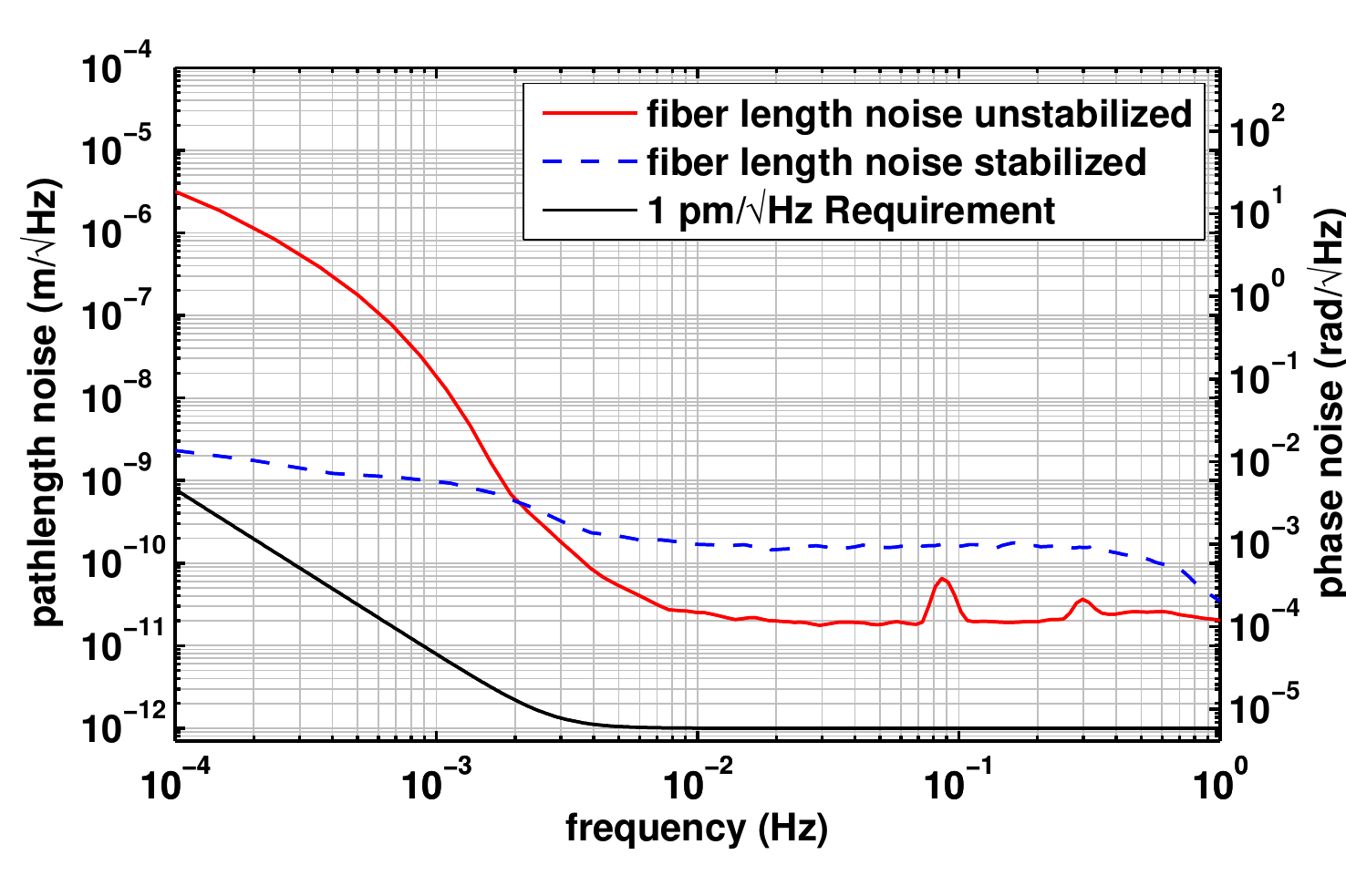}
	\caption[Fiber length noise with/without active length stabilization]{Comparison between fiber length noise with and without active length stabilization.}
	\label{fig:fiber_length_stab_vs_unstab}
\end{figure}

The effect of fiber length stabilization on fiber length noise is shown in Figure \ref{fig:fiber_length_stab_vs_unstab}. The solid red trace shows the noise spectral density of fiber length noise encountered in a typical measurement. The noise shape is flat at a level of about 20 pm/\rootHz{} for frequencies above 10\,mHz but rises proportional to 1/$f^2$ toward lower frequencies. With the fiber length stabilization enabled, as represented by the dashed blue trace, this noise level increases to 200 picometer/\rootHz{} for frequencies above 2\,mHz, probably due to voltage noise of the high voltage amplifier used to generate the piezo drive voltage. Taking into account that the performance achieved with this stabilization was sufficient to reduce non-reciprocal noise levels to the picometer level and that despite this decrease in length stability at high frequencies the non-reciprocal noise in this frequency range did not increase, no effort was made to further investigate the origin of the decreased length stability at high frequencies. At low frequencies, the length noise level does not increase much further toward even lower frequencies, such that one observes a noise level of only 2 nm/\rootHz{} at 0.1\,mHz: much lower than the 2 \textmu m/\rootHz{} measured without the length stabilization.

\begin{figure}
	\centering
		\includegraphics[width=\columnwidth]{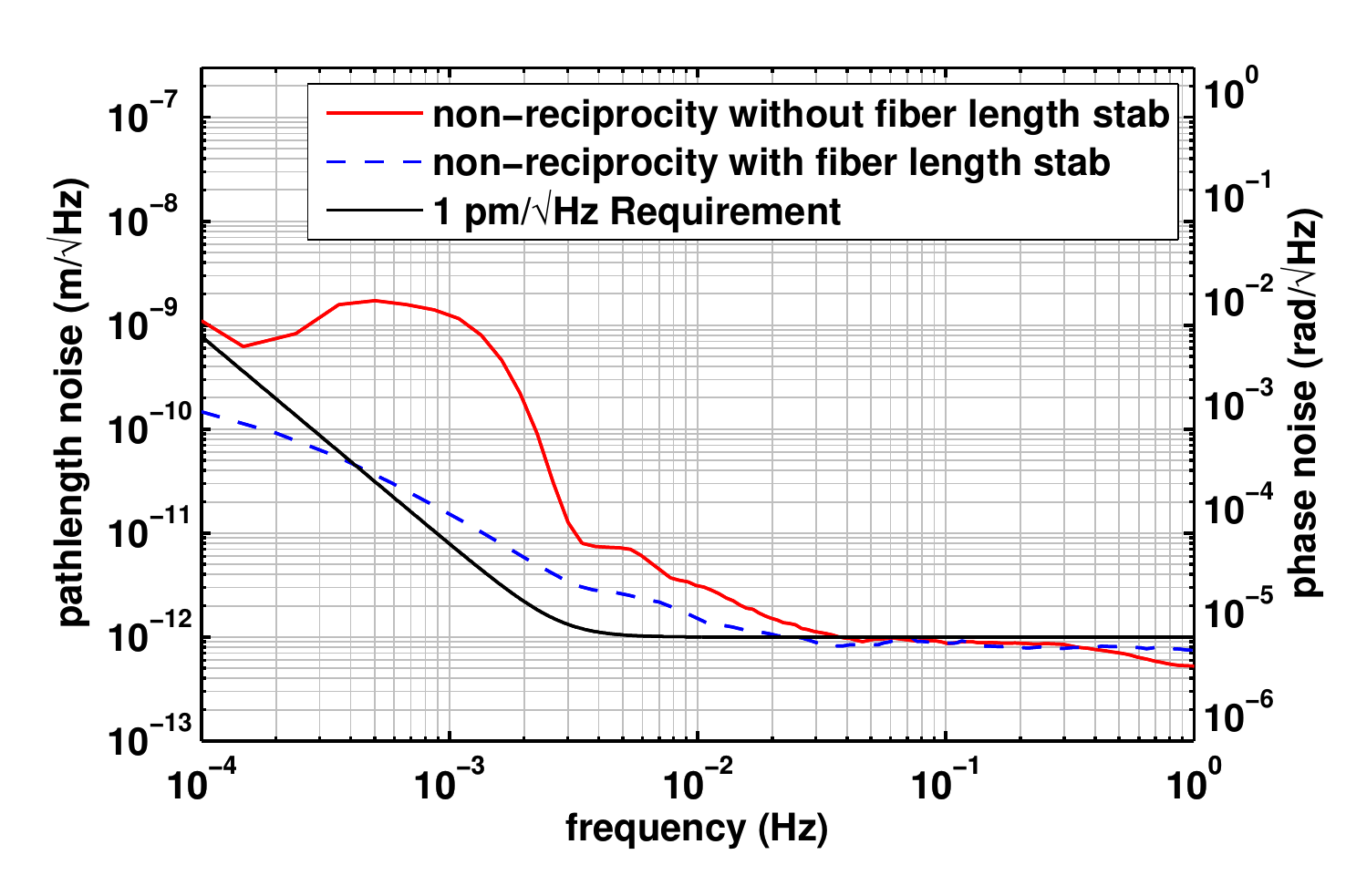}
	\caption[Non-reciprocal noise observed with/without fiber length stabilization]{Comparison of non-reciprocal noise observed without fiber length stabilization to noise observed with active fiber length stabilization.}
	\label{fig:compare_no_fib_quiet_normal}
\end{figure}

The effect on the measurement of the non-reciprocal fiber noise is illustrated in Figure \ref{fig:compare_no_fib_quiet_normal}. The red trace shows the non-reciprocal noise spectral density in a measurement where the fiber length stabilization was not active. While the noise is at the 1\,pm/\rootHz{} level for frequencies above 30\,mHz, it rises like $1/f$ for frequencies between 3\,mHz and 30\,mHz and a large ``bump'' with a peak height of 1\,nm/\rootHz{} is observed around 1\,mHz. Comparing this to the dahed blue trace, representing the noise observed with active fiber length stabilization, one sees that the active fiber length stabilization removes the bump observed around 1\,mHz and leads to an observed non-reciprocal noise level very close to the 1\,pm/\rootHz{} requirement represented by the solid black trace.

\section{Fiber non-reciprocity and noise subtraction in data post-processing}

\begin{figure*}
	\centering
		\includegraphics[width=\textwidth]{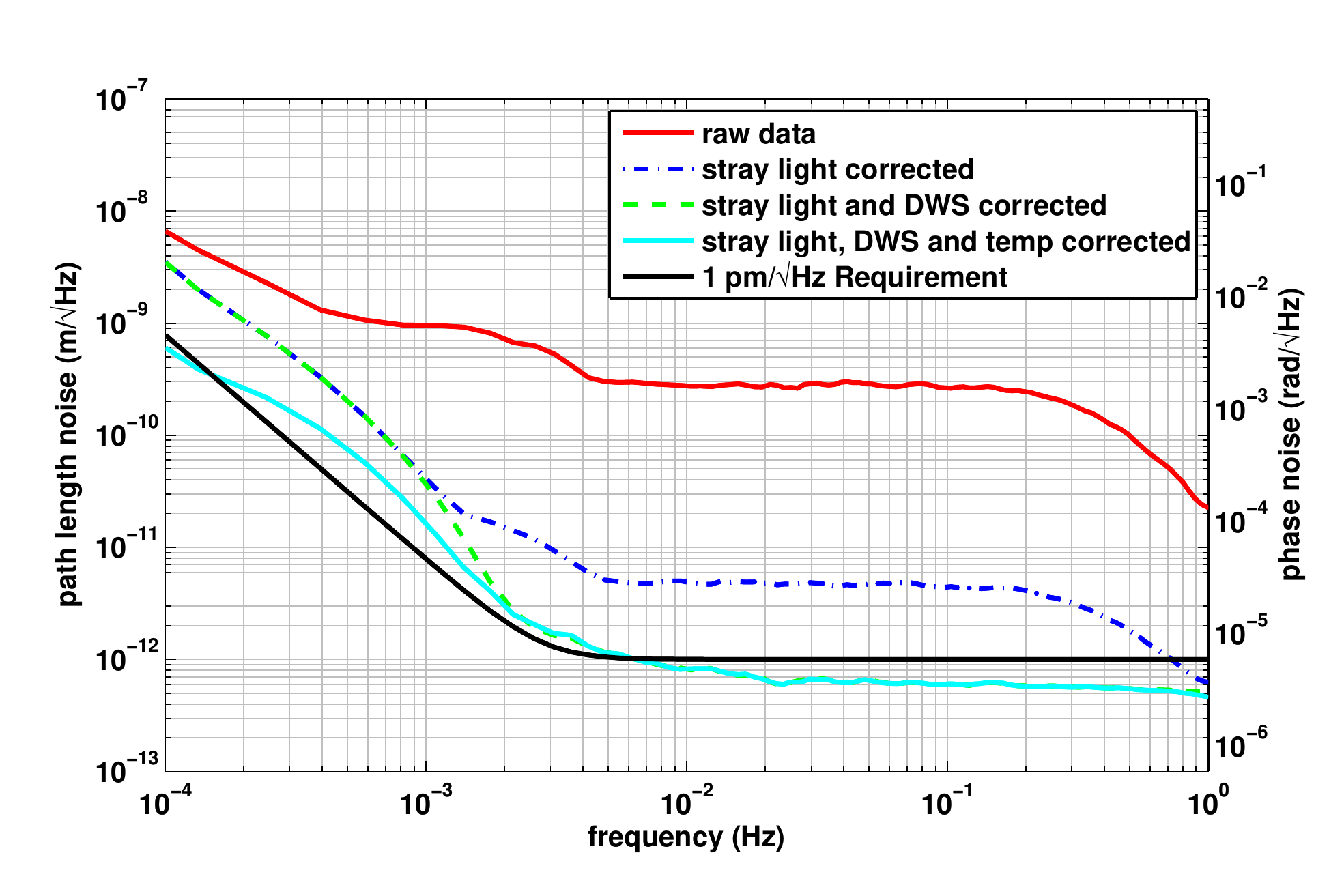}
	\caption{Non-reciprocity measurements using a single-mode polarization maintaining fiber, with corrections applied}
	\label{fig:non-rec-fiber-with-corrections}
\end{figure*}

Even after identifying and eliminating external influences from the measurement system, as discussed above, the observed non-reciprocal noise level was still higher than the requirements. The raw non-reciprocal noise found is represented by the red trace in Figure \ref{fig:non-rec-fiber-with-corrections}. To further improve the measurement sensitivity, a series of noise subtraction techniques was applied to the data acquired. While it is in principle possible to do so in real-time, the subtractions were applied in data post processing here.

Among the sources of noise that we identified and subtracted was scattered light. This light stems from Rayleigh backscatter in the fiber under test \cite{Barnoski77} as well as light scattered at the fiber coupler assemblies and their associated interfaces. This back-reflected light cannot be blocked because it travels along the same beam axis and is in the same spatial mode as the counter propagating beam when it originates from a reflection occurring inside the fiber.

\begin{figure*}
	\centering
		\includegraphics[width=\textwidth]{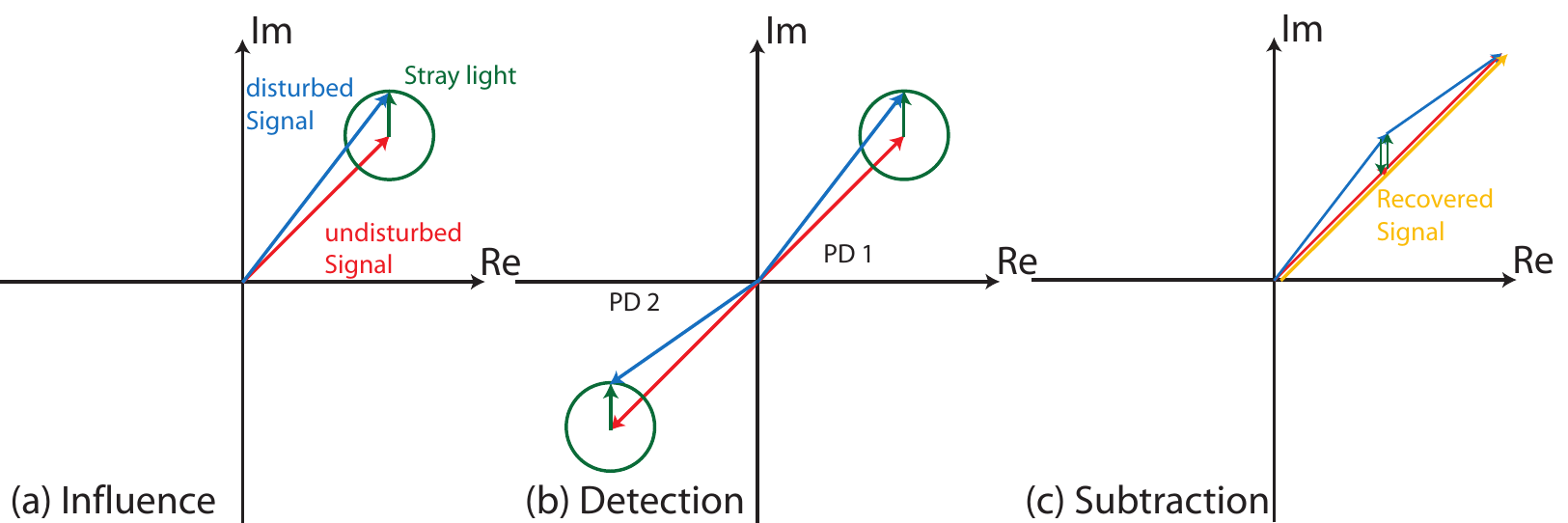}
		\caption{Illustration showing the principle of stray light subtraction in balanced detection}
	\label{fig:StraylightCombined}
\end{figure*}

Therefore, a balanced detection \cite{Painchaud2009} scheme was applied. We used photo detectors in both output ports of the recombination beam splitters and recorded their phases and amplitudes individually. Using a combination of these two signals the spurious signal stemming from the reflected light can be suppressed while the desired beat note is amplified. This is illustrated in Figure \ref{fig:StraylightCombined}. Nominally, the beat notes in both outputs of the beam splitter differ by 180\degree{} in phase. Therefore, subtracting the two signals from each other yields a signal of twice the amplitude. However, stray light signals will be in phase in both outputs if they did not originate from an interference occurring at the beam splitter. This leads to a common mode reduction of the stray light signal when the difference of both output channels is computed.

As data was readily available in the form of amplitude and phase of the signals, subtraction was done by calculating the vector sum of one phase vector and the inverted other phase vector.

However, it was found that great care had to be taken in order to achieve optimal suppression of the spurious signal. For such an optimal subtraction it is mandatory that the two measurement signals be of exactly equal amplitude. As can be seen in the same figure, a mismatch in amplitudes leads to a less effective subtraction of spurious signals.

To overcome the limitations of the direct stray light subtraction in the case of unmatched signal amplitudes, a normalized stray light correction was implemented, where the signal amplitude was taken into account for the correction. The two signals that were to be combined were divided by their amplitude first, thereby allowing for a much better suppression of the unwanted stray light signal. This is described in detail in \cite{Fleddermann2011JPCS}.

The resulting noise spectral density after this subtraction is shown in Figure \ref{fig:non-rec-fiber-with-corrections} in blue. Comparing this trace to the original, unmodified data represented by the red trace reveals that the subtraction led to a reduction of observed non-reciprocal noise by about a factor of 60 in the frequency range between 1~mHz and 0.1~Hz.

\subsubsection{Beam pointing fluctuations}

Beam pointing fluctuations can potentially lead to spurious changes in the measured path length, e.g. when beams pass through wedged optics, where the optical path length depends on the position at which the beam passes through. It was therefore decided to use the data available from quadrant photo detectors (QPDs) to measure these beam pointing fluctuations using a combination of the signals from the four channels of the detector, using a technique known as differential wavefront sensing (DWS). This allows for very sensitive measurement of the relative tilt of the two beams impinging onto the photo diode as is described in detail in \cite{Morrison1994,Morrison2:94}.

\begin{figure}
	\centering
		\includegraphics[width=\linewidth]{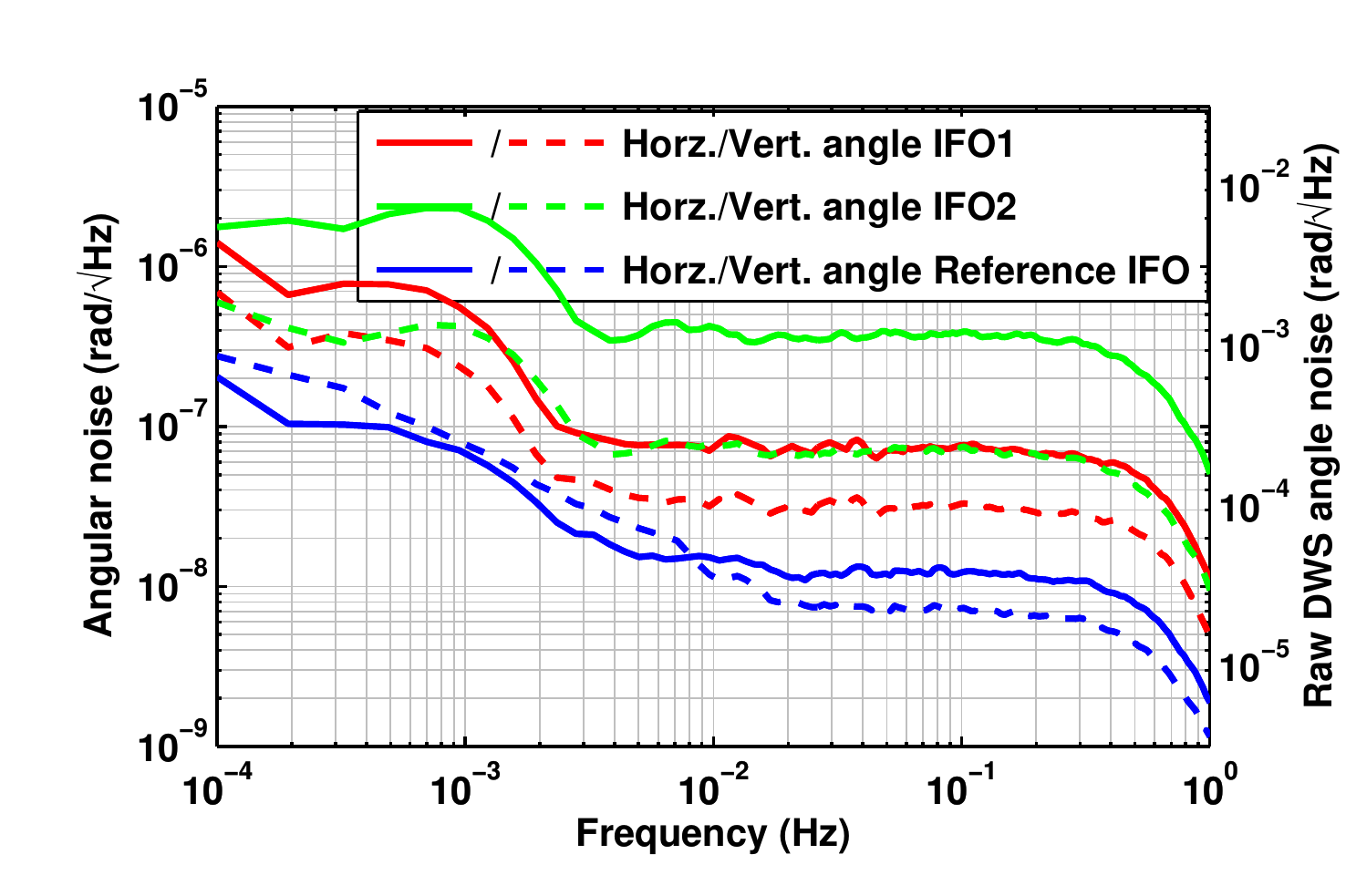}
	\caption{Typical beam pointing fluctuations during the measurement}
	\label{fig:DWS_noise_paper2}
\end{figure}

When comparing the shape of the noise curve after stray light subtraction in Figure \ref{fig:non-rec-fiber-with-corrections} to the shape of the angular noise found in Figure \ref{fig:DWS_noise_paper2}
one can see that they show a similar frequency dependence at frequencies above 5\,mHz. It was therefore concluded that beam pointing may couple to non-reciprocal path length changes. The reason this can lead to non-reciprocal noise is that pointing fluctuations at both fiber ends are independent, and therefore the effects are not necessarily identical on the light having traveled through the fiber in opposite directions.

The subtraction was done through a linear fit after low pass filtering the data, very similar to the process described in \cite{Guzman2008APB}. The resulting noise spectral density after this subtraction is again shown in Figure \ref{fig:non-rec-fiber-with-corrections}. Comparing the green trace representing the data after both stray light subtraction and beam pointing subtraction had been applied to the previous one, shown in blue, where only stray light subtraction had been applied, reveals that this additional subtraction led to a reduction of observed non-reciprocal noise by about another factor of 8 in the frequency range between 1~mHz and 0.1~Hz.

\subsubsection{Temperature noise subtraction}

\begin{figure}
	\centering
		\includegraphics[width=\columnwidth]{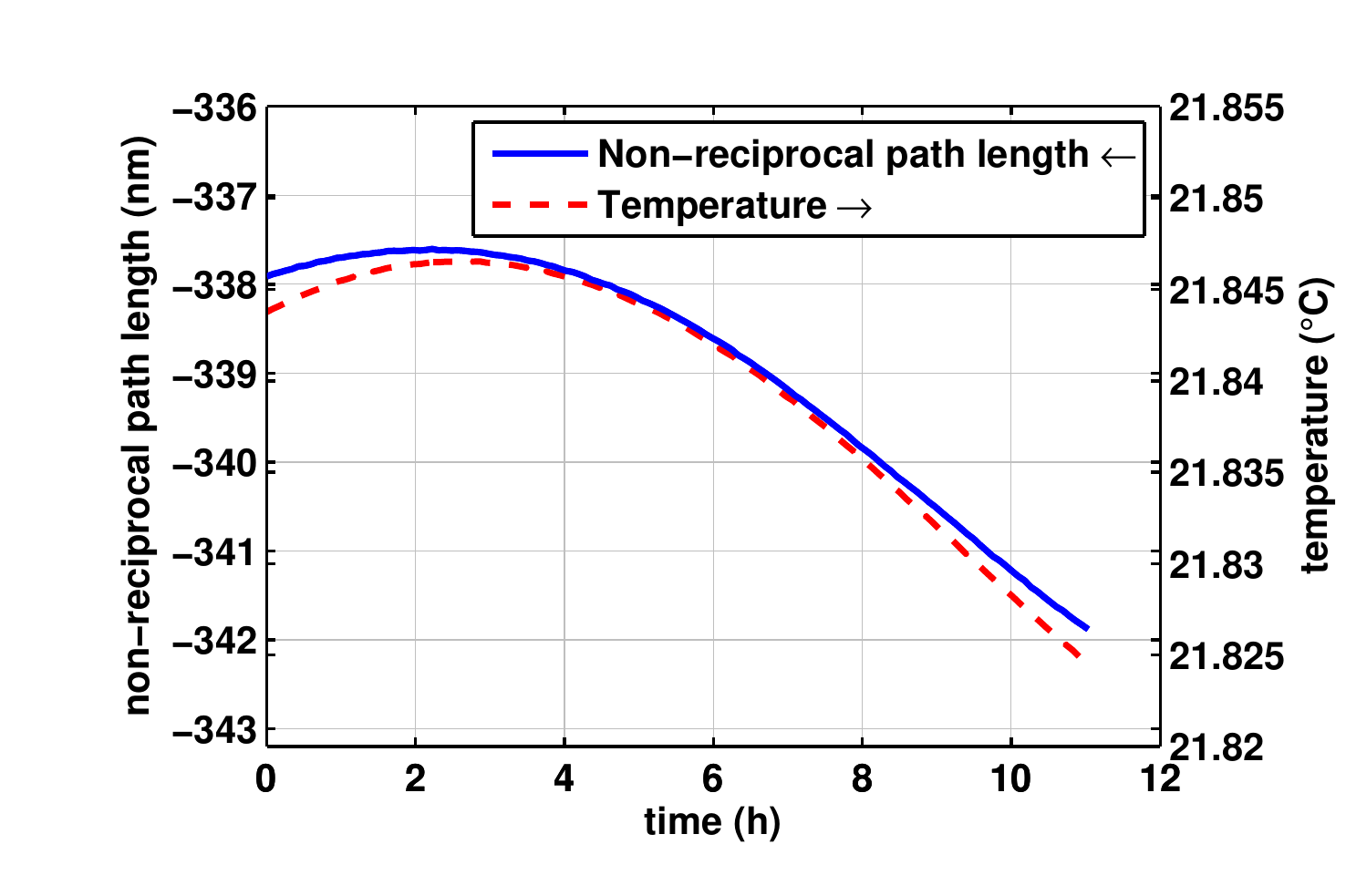}
	\caption{Observed correlations between temperature and non-reciprocal phase}
	\label{fig:temp_corr_fiber}
\end{figure}

Despite the low coefficient of thermal expansion of the Zerodur\texttrademark{} base plate, it was still found that for very low frequencies (below 1~mHz) a correlation between the temperature of the base plate and the non-reciprocal phase could be observed, as can be seen in Figure \ref{fig:temp_corr_fiber}. In this figure, the blue trace represents the non-reciprocal phase noise, plotted on the left hand y-axis compared to the temperature of the Zerodur\texttrademark{} base plate, plotted on the right hand y-axis to allow for the different units and scale. As one can see, these two lines follow very similar shapes during the course of about 11\,hours during which the data was recorded. One potential reason for this is non-uniform thermal expansion of the base plate, leading to different phase changes in the two measurement interferometers through  a change in length of a path that is only traversed by one of the beams. Another candidate explanation is the temperature dependence of the optic's refractive index. The change of refractive index with temperature of fused silica is typically 1.28$\times 10^{-5}$/K, about two orders of magnitude higher than the physical thermal expansion of the base plate. While the non-common path length in the medium was kept as short as possible, this still led to a residual coupling.

To remove this influence, the temperature of the Zerodur\texttrademark{} base plate was recorded using a digital thermometer. The obtained data was then used to subtract the effect of thermal expansion from the measurements. Again, data was low pass filtered and a linear fit was used to find the coefficient. The resulting noise spectral density after this subtraction is again shown in Figure \ref{fig:non-rec-fiber-with-corrections}. 

Comparing the cyan trace representing the data after all three subtraction had been applied to the previous one, shown in green, where only stray light subtraction and beam pointing subtraction had been applied, reveals that this additional subtraction led to a reduction of observed non-reciprocal noise by about another factor of two in the frequency range below 1~mHz, representing the lowest observed non-reciprocal phase noise obtained using this setup. This noise is at a level of about 1~\,pm/\rootHz{} with an 1/$f^2$ increase towards frequencies below 2.8 mHz and therefore shows that the PANDA type polarization maintaining fiber used in this experiment meets the requirements for a possible fiber backlink for LISA.

\section{Conclusion}

We have experimentally demonstrated a fiber-based backlink for LISA within the requirement of 1\,pm/\sqrtHz{} using a polarization-maintaining single-mode step index fiber.

Spurious interference signals from back-scattered light in the fiber or back-reflected light from its interfaces required the use of balanced detection.
This has severe implications for the design of an optical bench for LISA. When balanced detection is used then four photo detectors per interferometer will be required for redundancy. 

We also measured the effects of beam angle and temperature changes on the fiber non-reciprocity and subtracted their effects. While the solution here is not as simple as originally envisaged it represents a feasible baseline design option. 

\section*{Acknowledgments}

%\ack
We acknowledge funding by the European Space Agency within the project `Optical Bench Development for LISA', and support by Deutsches
Zentrum f\"ur Luft und Raumfahrt (DLR) with funding from the Bundesministerium f\"ur Wirtschaft und Technologie (DLR project reference 50
OQ 0601). We thank the German Research Foundation for funding the cluster of Excellence QUEST - Centre for Quantum Engineering and Space-
Time Research.

\section*{References}
\bibliography{unpublished,library,refs}

\end{document}